\documentclass[twocolumn,aps,pre,reprint,amsmath]{revtex4-1}
\usepackage{graphics}

\begin{document}

\title{Mode-coupling theory predictions for the dynamical transitions
  of the partly pinned fluid systems}

\author{Vincent Krakoviack}
\affiliation{Laboratoire de Chimie, {\'E}cole Normale Sup{\'e}rieure
  de Lyon, 46 All{\'e}e d'Italie, 69364 Lyon cedex 07, France}

\date{\today}

\begin{abstract}
  The predictions of the mode-coupling theory (MCT) for the dynamical
  arrest scenarios in a partly pinned (PP) fluid system are
  reported. The corresponding dynamical phase diagram is found to be
  very similar to that of a related quenched-annealed (QA) system. The
  only significant qualitative difference lies in the shape of the
  diffusion-localization lines at high matrix densities, with a
  re-entry phenomenon for the PP system but not for the QA model, in
  full agreement with recent computer simulation results. This finding
  clearly lends support to the predictive power of the MCT for
  fluid-matrix systems. Finally, the predictions of the MCT are shown
  to be in stark contrast with those of the random first-order
  transition theory. The PP systems are thus confirmed as very
  promising models for tests of theories of the glass transition.
\end{abstract}

%\pacs{64.70.P-, 46.65.+g, 61.20.Lc}

\maketitle

A partly pinned (PP) fluid system is a model of a fluid in contact
with a disordered substrate which is obtained by instantaneously
arresting the motion of a fraction of the particles in an equilibrium
bulk system and letting the remaining mobile fraction evolve under the
influence of the static random environment generated by the pinned
particles. The pinning process can be homogeneous, if, for instance,
the pinned particles are randomly chosen in the whole volume of the
bulk system, or heterogeneous, if the arrested particles are those
located in a predefined region of space. In the first case, one gets a
model of a fluid adsorbed in a statistically homogeneous disordered
porous solid; in the second one, a model of a fluid confined by
amorphous rough walls.

Because of their peculiar preparation process from an equilibrium bulk
fluid in which the future mobile and pinned components influence each
other, these systems display a number of interesting features. For
instance, one can show that the realization-averaged configurational
properties of a PP system exactly match those of the bulk fluid from
which it is prepared \cite{SchKobBin04JPCB,Kra10PRE}. Stated
otherwise, the pinning step does not alter the average configurational
properties of the system. This obviously represents a major
simplification when studying these quenched-disordered models, which,
over the years, have been considered in various fields of liquid state
theory.

In the theory of adsorption in disordered porous solids, they appear
as special cases of fluids adsorbed in depleted or templated
matrices. Indeed, the first class corresponds to models of porous
solids obtained by freezing equilibrium configurations of a fluid and
removing at random a fraction of its particles
\cite{TasTalVioTar97PRE,Tas97JCP}, the second one to models obtained
by freezing equilibrium configurations of a binary mixture and
removing one component called the template
\cite{Tas99PRE,ZhaTas00JCP,ZhaTas00MP,ZhaCheTas01PRE}. Thus, it is
clear that a homogeneous PP system corresponds to a depleted or
templated fluid-matrix system in which the depleted or template
particles are reinjected as the confined fluid \cite{Kra10PRE}. Both
the depleted and templated models are extensions of the concept of
quenched-annealed (QA) system, first introduced by Madden and Glandt
\cite{MadGla88JSP}, in which the porous matrix is obtained by simply
quenching equilibrium configurations of a fluid without any subsequent
particle removal. The properties of these different models have been
compared by Van Tassel \textit{et al.}
\cite{TasTalVioTar97PRE,Tas97JCP,%
  Tas99PRE,ZhaTas00JCP,ZhaTas00MP,ZhaCheTas01PRE}. An interesting
finding that will be useful in the following is that, for a given
matrix density, the depleted and templated matrices generically show
stronger correlations and a more open accessible volume than a simple
quenched matrix \cite{TasTalVioTar97PRE,Tas97JCP,Tas99PRE}.

In computational studies of the dynamics of fluids in confinement have
also regularly appeared PP systems
\cite{VirAraMed95PRL,VirMedAra95PRE,SchKobBinPar02PMB,%
  SchKobBin02EL,SchKobBin04JPCB,Kim03EL,FenMryPryFol09PRE,ChaJagYet04PRE,%
  MitErrTru06PRE,ChaJuaMed08PRE,KimMiySai09EL,KimMiySai10EPJST,%
  KimMiySai11JPCM}. Here, the use of these models can turn out highly
advantageous. Indeed, the configurations of the fluid particles
obtained immediately after the pinning step are automatically
equilibrium configurations by construction
\cite{SchKobBin04JPCB,Kra10PRE}. Hence, an acceptable initial
configuration is always readily available, while finding one can be a
very nontrivial task for other models (see Ref.~\cite{KurCosKah10PRE}
for the case of QA systems), and it does not have to be equilibrated
\cite{SchKobBinPar02PMB,SchKobBin04JPCB}. This is clearly very
interesting for studies of dense and/or glass-forming systems.

Following insightful pioneering works where PP systems had been
investigated in order to measure the spatial extent of dynamical
correlations in confined glass-forming liquids
\cite{SchKobBinPar02PMB,SchKobBin02EL,SchKobBin04JPCB,Kim03EL}, it has
been realized that these systems could also be used to probe the
existence of nontrivial static correlations in bulk glassy
liquids. The key here is to consider the pinning process as a
constraint imposed on the fluid and to measure how the configurations
of the constrained particles influence the accessible states of the
free ones, via the computation of point-to-set correlations
\cite{BouBir04JCP,MonSem06JSP_2,MezMon06JSP,MonSem06JSP,FraMon07JPA},
for instance.  Numerous studies of PP systems have recently appeared
along this line \cite{CavGriVer07PRL,%
  BirBouCavGriVer08NatPhys,CavGriVer10JSMTE,CavGriVer10preprint,%
  KarPro11preprint,BerKob11preprint,CamBir11preprint,%
  KobRolBer11preprint,ChaChaTar11preprint}.

Finally, it might be worth mentioning that an experimental nearly
two-dimensional realization of such a system is possible, by simply
squeezing a binary colloidal mixture between two glass plates
\cite{CruSauAra98PRL,CruAra99PRE}.

In this paper, we contribute to the study of the glassy dynamics in PP
systems and report on the predictions of the mode-coupling theory
(MCT) \cite{gotzebook} for the dynamical arrest scenarios in the
homogeneous case. To this end, we use a recent extension of the MCT to
fluids imbibed in disordered porous solids
\cite{Kra05PRL,Kra05JPCM,Kra07PRE,Kra09PRE} and compute the dynamical
phase diagram of the pinned system deriving from the one-component
hard-sphere fluid, which, thanks to the overall mathematical structure
of the theory and the robustness of the associated bifurcation schemes
\cite{gotzebook}, can be expected to be a representative example.

The motivation for this work is twofold. First, computer simulation
studies of the slow dynamics of fluids confined in disordered porous
matrices have recently appeared \cite{KurCosKah09PRL,KurCosKah10PRE,%
  KurCosKah11JPCM,KimMiySai09EL,KimMiySai10EPJST,KimMiySai11JPCM},
allowing comparisons with the previous predictions of the MCT
\cite{Kra05PRL,Kra05JPCM,Kra07PRE,Kra09PRE}. In many respects, they
demonstrate that the MCT offers a consistent, though idealized,
picture of the dynamics of these systems. For instance, the observed
changes in the density fluctuation relaxation pattern with increasing
confinement agree well with the theoretical picture of a crossover
from type B (bulk-like) bifurcation scenarios at low matrix densities
to type A (Lorentz-gas-like) scenarios at high matrix
densities. However, a potential issue with the theory is also pointed
out. Indeed, while the MCT predicts a dynamical re-entry phenomenon at
high matrix densities for the equisized hard-sphere QA system, no sign
of it is actually visible in the simulation data
\cite{KurCosKah09PRL,KurCosKah10PRE,%
  KurCosKah11JPCM,KimMiySai09EL,KimMiySai10EPJST,KimMiySai11JPCM}. But,
interestingly, Kim \emph{et al.} do clearly find such a re-entry
phenomenon in their results for the PP variant of this system
\cite{KimMiySai09EL,KimMiySai10EPJST,KimMiySai11JPCM}. One aim of the
present work is to show, after some clarifications, that the MCT is in
complete agreement with the results of the computer simulations with
respect to this re-entry phenomenon at high matrix densities, even
with the seemingly negative result obtained for the hard-sphere QA
system.

Second, the effect of pinned particles on the glass transition has
recently been investigated in the framework of the random first-order
transition (RFOT) theory \cite{KirThiWol89PRA,LubWol07ARPC} by
Cammarota and Biroli, who find a very interesting and nontrivial
scenario \cite{CamBir11preprint}. It is widely believed that the RFOT
theory and the MCT have strong connections, because they have some
mathematical structures in common. However, actual calculations
sometimes reveal major discrepancies, like when the spatial dimension
is changed \cite{SchSch10PRE,IkeMiy10PRL,ChaIkeParZam11preprint}. It
thus seems very natural to wonder how the two theories compare in the
present context. As we will show, it turns out that the scenario
predicted by the MCT differs very significantly from the one obtained
by Cammarota and Biroli.

The application of the MCT scheme to fluid-matrix systems results in
self-consistent equations for the time evolution of $\phi_q(t)$, the
normalized connected autocorrelation function of the fluid density
fluctuations, and $\phi^\text{s}_q(t)$, the autocorrelation function
of the tagged-particle density fluctuations, at wave vector modulus
$q$ \cite{Kra07PRE,Kra09PRE}.  In the infinite time limit, from which
the state of the system can be determined, they reduce to equations
for the nonergodicity parameters $f_q= \lim_{t\to\infty} \phi_q(t)$
and $f^\text{s}_q= \lim_{t\to\infty} \phi^\text{s}_q(t)$, which read
\begin{gather}
  \frac{f_q}{1-f_q} = \int \frac{d^3\mathbf{k}}{(2\pi)^3} \left[
    V^{(2)}_{\mathbf{q},\mathbf{k}} f_{k} f_{|\mathbf{q-k}|} +
    V^{(1)}_{\mathbf{q},\mathbf{k}} f_{k} \right], \label{eq1} \\
  \frac{f^\text{s}_q}{1-f^\text{s}_q} = \int
  \frac{d^3\mathbf{k}}{(2\pi)^3} \left[
    v^{(2)}_{\mathbf{q},\mathbf{k}} f^\text{s}_{k} f_{|\mathbf{q-k}|}
    + v^{(1)}_{\mathbf{q},\mathbf{k}} f^\text{s}_{k}
  \right], \label{eq2}\\
  V^{(2)}_{\mathbf{q},\mathbf{k}} = \frac{1}{2} n_\text{f}
  S^\text{c}_q \left[\frac{\mathbf{q}\cdot\mathbf{k}}{q^2}
    \hat{c}^\text{c}_k + \frac{\mathbf{q}\cdot(\mathbf{q-k})}{q^2}
    \hat{c}^\text{c}_{|\mathbf{q-k}|}\right]^2 S^\text{c}_k
  S^\text{c}_{|\mathbf{q-k}|},\\
  V^{(1)}_{\mathbf{q},\mathbf{k}} = n_\text{f} S^\text{c}_q \left[
    \frac{\mathbf{q}\cdot\mathbf{k}}{q^2} \hat{c}^\text{c}_k +
    \frac{\mathbf{q}\cdot(\mathbf{q-k})}{q^2} \frac{1}{n_\text{f}}
  \right]^2 S^\text{c}_k S^\text{d}_{|\mathbf{q-k}|},\\
  v^{(2)}_{\mathbf{q},\mathbf{k}} = n_\text{f} \left[
    \frac{\mathbf{q}\cdot(\mathbf{q-k})}{q^2} \right]^2 \left[
    \hat{c}^\text{c}_{|\mathbf{q-k}|} \right]^2
  S^\text{c}_{|\mathbf{q-k}|},\\
  v^{(1)}_{\mathbf{q},\mathbf{k}} = \left[
    \frac{\mathbf{q}\cdot(\mathbf{q-k})}{q^2} \right]^2
  \hat{h}^\text{d}_{|\mathbf{q-k}|},
\end{gather} 
where $n_\text{f}$ is the fluid density, $S^\text{c}_q$ the connected
structure factor, $S^\text{d}_q$ the disconnected structure factor,
$\hat{h}^\text{d}_{q}$ the Fourier transform of the disconnected total
pair correlation function, and $\hat{c}^\text{c}_q$ the Fourier
transform of the connected direct correlation function defined as $
n_f \hat{c}^{c}_q = 1 - 1/S^{c}_q$
\cite{GivSte92JCP,LomGivSteWeiLev93PRE,GivSte94PA,RosTarSte94JCP}.
 
While Refs.~\cite{Kra05PRL,Kra05JPCM,Kra07PRE,Kra09PRE} explicitly
mention QA systems, no assumption about the statistics of the
disordered solid is actually made during the derivation of the MCT. It
follows that it can be applied to any fluid-matrix model, and in
particular to homogeneous PP systems. Such a possibility has sometimes
been challenged \cite{ChaJagYet04PRE,SchKobBin04JPCB}. Indeed, as
already mentioned, the pinning protocol exactly preserves the
configurational properties of the bulk fluid from which a PP system is
prepared \cite{SchKobBin04JPCB,Kra10PRE}, but it can have a strong
effect on the dynamics \cite{VirAraMed95PRL,%
  VirMedAra95PRE,SchKobBinPar02PMB,SchKobBin02EL,SchKobBin04JPCB,%
  Kim03EL,FenMryPryFol09PRE,ChaJagYet04PRE,MitErrTru06PRE,%
  ChaJuaMed08PRE,KimMiySai09EL,KimMiySai10EPJST,KimMiySai11JPCM}. At
first sight, this looks incompatible with the use of the MCT, which is
known to take only structural quantities as input in order to predict
the dynamics. The present worked example should clearly show that no
such difficulty actually exists, the key being that, for fluid-matrix
systems, one has to properly take into account the fact that the
correlation functions generically split into connected and
disconnected components in the presence of quenched disorder, due to
broken symmetries at the microscopic level \cite{LifGrePasbook}. In
other words, even if this is not visible at the level of simple
configurational properties, the structure does actually change in the
pinned system (see the Supplementary Material of
Ref.~\cite{KobRolBer11preprint} for an illustration). Similar care
should be taken when discussing the thermodynamics of PP systems,
since the thermodynamic susceptibilities such as the heat capacity and
the compressibility are given by connected correlations in disordered
systems \cite{ForGla94JCP,RosTarSte94JCP}.

In addition to the results for the equisized hard-sphere PP system,
those for the analogous QA system, which already appeared in
Refs.~\cite{Kra05PRL,Kra05JPCM,Kra07PRE,Kra09PRE}, will be recalled
for comparison. The difference between the two models is subtle. In
both cases, the fluid and the matrix consist of hard-sphere particles
of the same diameter and the control parameters are the volume
fractions occupied by the fluid and matrix particles, denoted by
$\phi_\mathrm{f}$ and $\phi_\mathrm{m}$, respectively. Only the
preparation protocol is slightly modified. In the QA system, the
configurations of the matrix particles are drawn from those of an
equilibrated hard-sphere fluid of compacity $\phi_\mathrm{m}$, then a
fluid of mobile particles with compacity $\phi_\mathrm{f}$ is inserted
in the obtained disordered samples. In the PP system, a hard-sphere
fluid of compacity $\phi_\mathrm{f}+\phi_\mathrm{m}$ is first
equilibrated, then a fraction $x = \phi_\mathrm{m} /
(\phi_\mathrm{f}+\phi_\mathrm{m})$ of the particles is pinned down to
form the porous matrix of compacity $\phi_\mathrm{m}$, while the
remaining ones become the mobile fluid component with compacity
$\phi_\mathrm{f}$. So, while the matrix in the QA system is prepared
independently of the fluid that will be adsorbed in it, the matrix in
the PP system is prepared under its direct influence. This is the
reason for the generically stronger correlations and more open
accessible volume of the matrix in the PP system
\cite{TasTalVioTar97PRE,Tas97JCP,Tas99PRE}. Note that by construction,
both models coincide when $\phi_\mathrm{f}$ or $\phi_\mathrm{m}$
vanish.

The required structural input is computed for both systems within the
Percus-Yevick approximation
\cite{GivSte92JCP,LomGivSteWeiLev93PRE,GivSte94PA,MerLevWei96JCP},
which leads to analytic expressions for the hard-sphere PP system, as
it was recently realized \cite{Kra10PRE}. The numerical procedure used
to solve the above equations is described in
Refs.~\cite{Kra07PRE,Kra09PRE}, to which the interested reader is
referred for technical details.

\begin{figure}
  \includegraphics{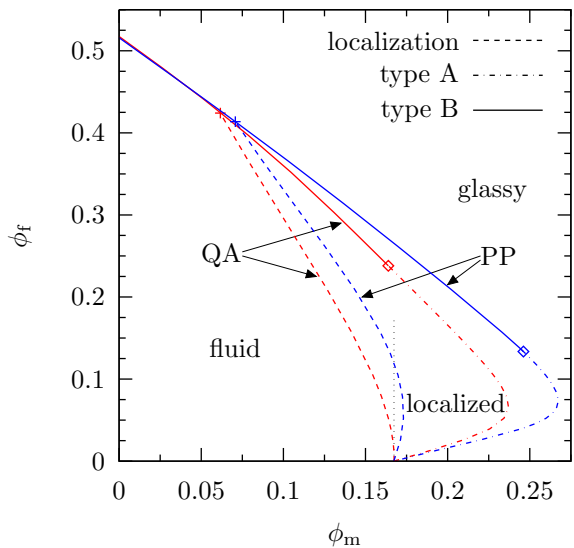}
  \caption{Mode-coupling dynamical phase diagrams of the equisized
    hard-sphere quenched-annealed (QA) and partly pinned (PP)
    fluid-matrix models. $\phi_\mathrm{f}$ and $\phi_\mathrm{m}$
    denote the volume fractions occupied by the fluid and matrix
    particles, respectively. The dotted vertical line is a guide to
    emphasize the re-entrant behavior of the diffusion-localization
    line of the PP system.}
  \label{fig.1}
\end{figure}

The resulting phase diagrams are plotted in Fig.~\ref{fig.1}. Both
have much in common. Three phases are found: fluid when
$f_q=f^\text{s}_q=0$ is the only solution, localized when $f_q=0$ and
$f^\text{s}_q\neq0$, and glassy when $f_q\neq0$ and
$f^\text{s}_q\neq0$. The systems enter the glassy phase by crossing an
ideal glass transition line, which corresponds to the freezing of the
collective dynamics and along which the bifurcation scenario changes
from type B to type A as the confinement gets stronger. Furthermore,
this line is preceded at high enough matrix densities by a
diffusion-localization transition line associated with a continuous
arrest of the self dynamics only. For the PP system, there might be a
very short glass-glass transition line at the junction between the
type A and B glass transition branches, but given its very limited
extent and the possible sources of inaccuracies in the numerical
calculations, such as the discretization of the integrals in
Eqs.~\eqref{eq1} and \eqref{eq2}, no definite conclusion can be
reached at this point. In practice, both systems can be said to have
phase diagrams with the same topology.

Now considering the shape of the phase diagrams, another shared
feature of both models is a significantly re-entrant ideal glass
transition line, the effect being more pronounced for the PP
system. But, and quite remarkably, the MCT delivers contrasting
predictions with respect to the behavior of the diffusion-localization
transition lines. Indeed, no re-entry phenomenon is visible in
Fig.~\ref{fig.1} for the QA system, while one is clearly present for
the PP system \footnote{In fact, the MCT does predict a re-entrant
  behavior for the diffusion-localization transition line of the QA
  system as well, but with an exceedingly small amplitude. The
  difficulties raised by this result and a possible physical
  interpretation are discussed in Ref.~\cite{Kra09PRE}.}. For both
types of transition, the overall trend, which is that the
ergodicity-breaking events occur at higher densities in the PP system,
is perfectly in line with the fact that it is characterized by a more
open accessible volume \cite{TasTalVioTar97PRE,Tas97JCP,Tas99PRE},
thus confirming an argument by Kim \textit{et al.}
\cite{KimMiySai09EL,KimMiySai10EPJST,KimMiySai11JPCM}.

The diffusion-localization transition lines are precisely those
investigated by Kim \textit{et al.} in their simulations
\cite{KimMiySai09EL,KimMiySai10EPJST,KimMiySai11JPCM}. Indeed, the
criterion they use to define the dynamically arrested phase is based
on the mean-squared displacement, a self dynamical quantity. This is
confirmed by the careful analysis of the interplay between the
collective and self dynamics led by Kurzidim \emph{et al.}
\cite{KurCosKah10PRE,KurCosKah11JPCM}. Hence, the conclusion: the MCT
does correctly capture the fact that the subtle structural differences
between the QA and PP models result in different re-entry behaviors of
their \emph{self} dynamics in the high matrix density regime, in full
agreement with the simulation results
\cite{KimMiySai09EL,KimMiySai10EPJST,KimMiySai11JPCM}.

Unfortunately, the situation is less clear with respect to the
\emph{collective} dynamics, which has been studied in detail for the
QA system only \cite{KurCosKah09PRL,KurCosKah10PRE}. Indeed, in
qualitative agreement with the prediction of distinct transition lines
for the self and collective dynamics at moderate and high matrix
densities, a wide separation of time scales between the two dynamics
is observed in simulations, but with no sign of a re-entrant
collective dynamics. The problem might lie in the theory or in the
simulations. Indeed, on the one hand, it is possible that the standard
formulation of the MCT based on equilibrium quantities breaks down if
the particles cannot redistribute themselves across the system, as it
is precisely the case in a localized state. On the other hand, the
simulations, which are very difficult in this density regime, do not
seem consistent with the expectation that the collective and self
dynamics coincide in the limit of vanishing fluid density. So, more
work is definitely needed to clarify this point. Here, the PP systems
could be very useful, thanks to the possibility to easily generate
equilibrated samples even at high densities.

\begin{figure}
\includegraphics{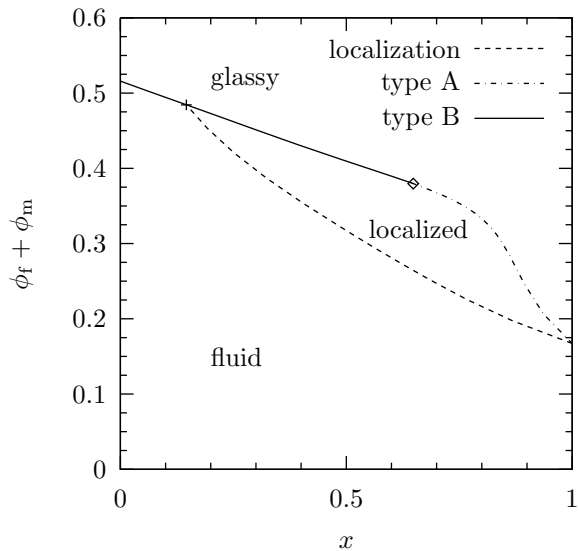}
\caption{Mode-coupling dynamical phase diagram of the equisized
  hard-sphere partly pinned fluid-matrix
  model. $\phi_\mathrm{f}+\phi_\mathrm{m}$ is the total volume
  fraction occupied by both the fluid and matrix particles, $x =
  \phi_\mathrm{m} / (\phi_\mathrm{f}+\phi_\mathrm{m})$ is the pinning
  fraction.}
\label{fig.2}
\end{figure}

We now turn to the comparison with the predictions of the RFOT theory
\cite{CamBir11preprint}. For this, it is more convenient to consider
the dynamical phase diagram of the PP system in the plane defined by
the total volume fraction $\phi_\mathrm{f}+\phi_\mathrm{m}$ and the
pinning fraction $x$, as in Fig.~\ref{fig.2}. In passing, we note
that, thanks to this representation, the total densities at the
various transitions are clearly shown to be decreasing functions of
$x$, in line with the rather natural expectation that pinning down
particles slows down the dynamics.

Within the RFOT theory, the PP systems are found to occupy a very
special place. Indeed, at different levels of calculation (mean-field
or real-space renormalization group analysis), the theory consistently
predicts ideal glass transition lines starting at $x=0$ (bulk limit)
and ending at some critical value $x_\text{c}<1$. This means that one
can go from the fluid to the glassy phases without crossing any
transition, simply by moving around this terminal point as one can do
with the critical point of the liquid-gas transition. However, this is
an isolated case. For other types of random environments, it is
predicted that, if the disorder becomes strong enough, the glass
transition does not disappear, but turns continuous, very much like in
the MCT with the type A bifurcations, and it is then impossible to
reach the glassy phase without crossing a transition.

So, while the predictions of the MCT and the RFOT theory tend to agree
for generic cases of fluids in disordered environments, they are in
stark contrast for the PP systems. Indeed, within the MCT, the PP
systems appear as rather ordinary fluid-matrix systems, as illustrated
by Fig.~\ref{fig.1}, and the transition lines run up to $x=1$, so that
the system always has to cross a dynamical transition line in order to
enter the glassy phase, as shown in Fig.~\ref{fig.2}. The PP systems
are thus a clear example of diverging predictions between the two
considered theoretical schemes and, as such, would probably deserve
much attention in the future. We note that, based on the presently
available data and, in particular, on the joint study of QA and PP
systems by Kim \textit{et al.}
\cite{KimMiySai09EL,KimMiySai10EPJST,KimMiySai11JPCM}, no obvious
inconsistency with the MCT scenario has appeared yet, but, as pointed
out above, some issues are still pending and more work is needed.

In summary, in this paper, the predictions of the MCT for the slow
dynamics of a homogeneous PP system have been worked out and compared
to previous findings for a related QA system
\cite{Kra05PRL,Kra05JPCM,Kra07PRE,Kra09PRE}, to the results of recent
computer simulation studies
\cite{KimMiySai09EL,KimMiySai10EPJST,KimMiySai11JPCM} and to the
scenario derived within the framework of the RFOT theory
\cite{CamBir11preprint}. It is found that the dynamical phase diagrams
of the QA and PP systems are very similar within MCT. The only
significant qualitative difference is in the shape of the
diffusion-localization lines at high matrix densities, with a re-entry
phenomenon for the PP system but not for the QA model. In the light of
these results, the simulation data, which could first appear as a
challenge to the theory, are actually found to represent a nice
confirmation of its predictions. Finally, the comparison between the
MCT and the RFOT theory shows that the two approaches make predictions
that are in stark contrast.

The latter result might have serious consequences. Indeed, two
different points of view might be adopted on PP systems. On the one
hand, seen from the theory of simple liquids and adsorption phenomena,
they are fluid-matrix models among others \cite{TasTalVioTar97PRE,%
  Tas97JCP,Tas99PRE,ZhaTas00JCP,ZhaTas00MP,ZhaCheTas01PRE}. They
display interesting and possibly useful configurational properties
\cite{SchKobBin04JPCB,Kra10PRE}, but these are merely accidental and
no peculiar physics beyond confinement phenomena should be
expected. This is what the MCT supports. On the other hand, seen from
the theory of disordered systems, they are very special physical
objects, ideally and uniquely suited to track hidden forms of order
and the related phase transitions in amorphous systems
\cite{BouBir04JCP,MonSem06JSP_2,MezMon06JSP,%
  MonSem06JSP,FraMon07JPA,CavGriVer07PRL,BirBouCavGriVer08NatPhys,%
  CavGriVer10JSMTE,CavGriVer10preprint,KarPro11preprint,%
  BerKob11preprint,CamBir11preprint,KobRolBer11preprint,%
  ChaChaTar11preprint}. This is what the RFOT theory
illustrates. Finding which point of view is the most appropriate for
glass-forming liquids could have a profound influence on future
directions in glass transition theory.

\acknowledgments It is a pleasure to thank G. Biroli, C. Cammarota,
K. Kim, and K. Miyazaki for useful discussions.


%\bibliography{../../Bibtex/abbrev,../../Bibtex/confinedfluids_theo,%

\begin{thebibliography}{58}%
\makeatletter
\providecommand \@ifxundefined [1]{%
 \@ifx{#1\undefined}
}%
\providecommand \@ifnum [1]{%
 \ifnum #1\expandafter \@firstoftwo
 \else \expandafter \@secondoftwo
 \fi
}%
\providecommand \@ifx [1]{%
 \ifx #1\expandafter \@firstoftwo
 \else \expandafter \@secondoftwo
 \fi
}%
\providecommand \natexlab [1]{#1}%
\providecommand \enquote  [1]{``#1''}%
\providecommand \bibnamefont  [1]{#1}%
\providecommand \bibfnamefont [1]{#1}%
\providecommand \citenamefont [1]{#1}%
\providecommand \href@noop [0]{\@secondoftwo}%
\providecommand \href [0]{\begingroup \@sanitize@url \@href}%
\providecommand \@href[1]{\@@startlink{#1}\@@href}%
\providecommand \@@href[1]{\endgroup#1\@@endlink}%
\providecommand \@sanitize@url [0]{\catcode `\\12\catcode `\$12\catcode
  `\&12\catcode `\#12\catcode `\^12\catcode `\_12\catcode `\%12\relax}%
\providecommand \@@startlink[1]{}%
\providecommand \@@endlink[0]{}%
\providecommand \url  [0]{\begingroup\@sanitize@url \@url }%
\providecommand \@url [1]{\endgroup\@href {#1}{\urlprefix }}%
\providecommand \urlprefix  [0]{URL }%
\providecommand \Eprint [0]{\href }%
\providecommand \doibase [0]{http://dx.doi.org/}%
\providecommand \selectlanguage [0]{\@gobble}%
\providecommand \bibinfo  [0]{\@secondoftwo}%
\providecommand \bibfield  [0]{\@secondoftwo}%
\providecommand \translation [1]{[#1]}%
\providecommand \BibitemOpen [0]{}%
\providecommand \bibitemStop [0]{}%
\providecommand \bibitemNoStop [0]{.\EOS\space}%
\providecommand \EOS [0]{\spacefactor3000\relax}%
\providecommand \BibitemShut  [1]{\csname bibitem#1\endcsname}%
\let\auto@bib@innerbib\@empty
%</preamble>
\bibitem [{\citenamefont {Scheidler}\ \emph {et~al.}(2004)\citenamefont
  {Scheidler}, \citenamefont {Kob},\ and\ \citenamefont
  {Binder}}]{SchKobBin04JPCB}%
  \BibitemOpen
  \bibfield  {author} {\bibinfo {author} {\bibfnamefont {P.}~\bibnamefont
  {Scheidler}}, \bibinfo {author} {\bibfnamefont {W.}~\bibnamefont {Kob}}, \
  and\ \bibinfo {author} {\bibfnamefont {K.}~\bibnamefont {Binder}},\ }\href
  {\doibase 10.1021/jp036593s} {\bibfield  {journal} {\bibinfo  {journal} {J.
  Phys. Chem. B}\ }\textbf {\bibinfo {volume} {108}},\ \bibinfo {pages} {6673}
  (\bibinfo {year} {2004})}\BibitemShut {NoStop}%
\bibitem [{\citenamefont {Krakoviack}(2010)}]{Kra10PRE}%
  \BibitemOpen
  \bibfield  {author} {\bibinfo {author} {\bibfnamefont {V.}~\bibnamefont
  {Krakoviack}},\ }\href {\doibase 10.1103/PhysRevE.82.061501} {\bibfield
  {journal} {\bibinfo  {journal} {Phys. Rev. E}\ }\textbf {\bibinfo {volume}
  {82}},\ \bibinfo {pages} {061501} (\bibinfo {year} {2010})}\BibitemShut
  {NoStop}%
\bibitem [{\citenamefont {Van~Tassel}\ \emph {et~al.}(1997)\citenamefont
  {Van~Tassel}, \citenamefont {Talbot}, \citenamefont {Viot},\ and\
  \citenamefont {Tarjus}}]{TasTalVioTar97PRE}%
  \BibitemOpen
  \bibfield  {author} {\bibinfo {author} {\bibfnamefont {P.~R.}\ \bibnamefont
  {Van~Tassel}}, \bibinfo {author} {\bibfnamefont {J.}~\bibnamefont {Talbot}},
  \bibinfo {author} {\bibfnamefont {P.}~\bibnamefont {Viot}}, \ and\ \bibinfo
  {author} {\bibfnamefont {G.}~\bibnamefont {Tarjus}},\ }\href {\doibase
  10.1103/PhysRevE.56.R1299} {\bibfield  {journal} {\bibinfo  {journal} {Phys.
  Rev. E}\ }\textbf {\bibinfo {volume} {56}},\ \bibinfo {pages} {R1299}
  (\bibinfo {year} {1997})}\BibitemShut {NoStop}%
\bibitem [{\citenamefont {Van~Tassel}(1997)}]{Tas97JCP}%
  \BibitemOpen
  \bibfield  {author} {\bibinfo {author} {\bibfnamefont {P.~R.}\ \bibnamefont
  {Van~Tassel}},\ }\href {\doibase 10.1063/1.475249} {\bibfield  {journal}
  {\bibinfo  {journal} {J. Chem. Phys.}\ }\textbf {\bibinfo {volume} {107}},\
  \bibinfo {pages} {9530} (\bibinfo {year} {1997})}\BibitemShut {NoStop}%
\bibitem [{\citenamefont {Van~Tassel}(1999)}]{Tas99PRE}%
  \BibitemOpen
  \bibfield  {author} {\bibinfo {author} {\bibfnamefont {P.~R.}\ \bibnamefont
  {Van~Tassel}},\ }\href {\doibase 10.1103/PhysRevE.60.R25} {\bibfield
  {journal} {\bibinfo  {journal} {Phys. Rev. E}\ }\textbf {\bibinfo {volume}
  {60}},\ \bibinfo {pages} {R25} (\bibinfo {year} {1999})}\BibitemShut
  {NoStop}%
\bibitem [{\citenamefont {Zhang}\ and\ \citenamefont
  {Van~Tassel}(2000{\natexlab{a}})}]{ZhaTas00JCP}%
  \BibitemOpen
  \bibfield  {author} {\bibinfo {author} {\bibfnamefont {L.}~\bibnamefont
  {Zhang}}\ and\ \bibinfo {author} {\bibfnamefont {P.~R.}\ \bibnamefont
  {Van~Tassel}},\ }\href {\doibase 10.1063/1.480874} {\bibfield  {journal}
  {\bibinfo  {journal} {J. Chem. Phys.}\ }\textbf {\bibinfo {volume} {112}},\
  \bibinfo {pages} {3006} (\bibinfo {year} {2000}{\natexlab{a}})}\BibitemShut
  {NoStop}%
\bibitem [{\citenamefont {Zhang}\ and\ \citenamefont
  {Van~Tassel}(2000{\natexlab{b}})}]{ZhaTas00MP}%
  \BibitemOpen
  \bibfield  {author} {\bibinfo {author} {\bibfnamefont {L.}~\bibnamefont
  {Zhang}}\ and\ \bibinfo {author} {\bibfnamefont {P.~R.}\ \bibnamefont
  {Van~Tassel}},\ }\href {\doibase 10.1080/00268970009483357} {\bibfield
  {journal} {\bibinfo  {journal} {Mol. Phys.}\ }\textbf {\bibinfo {volume}
  {98}},\ \bibinfo {pages} {1521} (\bibinfo {year}
  {2000}{\natexlab{b}})}\BibitemShut {NoStop}%
\bibitem [{\citenamefont {Zhang}\ \emph {et~al.}(2001)\citenamefont {Zhang},
  \citenamefont {Cheng},\ and\ \citenamefont {Van~Tassel}}]{ZhaCheTas01PRE}%
  \BibitemOpen
  \bibfield  {author} {\bibinfo {author} {\bibfnamefont {L.}~\bibnamefont
  {Zhang}}, \bibinfo {author} {\bibfnamefont {S.}~\bibnamefont {Cheng}}, \ and\
  \bibinfo {author} {\bibfnamefont {P.~R.}\ \bibnamefont {Van~Tassel}},\ }\href
  {\doibase 10.1103/PhysRevE.64.042101} {\bibfield  {journal} {\bibinfo
  {journal} {Phys. Rev. E}\ }\textbf {\bibinfo {volume} {64}},\ \bibinfo
  {pages} {042101} (\bibinfo {year} {2001})}\BibitemShut {NoStop}%
\bibitem [{\citenamefont {Madden}\ and\ \citenamefont
  {Glandt}(1988)}]{MadGla88JSP}%
  \BibitemOpen
  \bibfield  {author} {\bibinfo {author} {\bibfnamefont {W.~G.}\ \bibnamefont
  {Madden}}\ and\ \bibinfo {author} {\bibfnamefont {E.~D.}\ \bibnamefont
  {Glandt}},\ }\href {\doibase 10.1007/BF01028471} {\bibfield  {journal}
  {\bibinfo  {journal} {J. Stat. Phys.}\ }\textbf {\bibinfo {volume} {51}},\
  \bibinfo {pages} {537} (\bibinfo {year} {1988})}\BibitemShut {NoStop}%
\bibitem [{\citenamefont {Viramontes-Gamboa}\ \emph
  {et~al.}(1995{\natexlab{a}})\citenamefont {Viramontes-Gamboa}, \citenamefont
  {Arauz-Lara},\ and\ \citenamefont {Medina-Noyola}}]{VirAraMed95PRL}%
  \BibitemOpen
  \bibfield  {author} {\bibinfo {author} {\bibfnamefont {G.}~\bibnamefont
  {Viramontes-Gamboa}}, \bibinfo {author} {\bibfnamefont {J.~L.}\ \bibnamefont
  {Arauz-Lara}}, \ and\ \bibinfo {author} {\bibfnamefont {M.}~\bibnamefont
  {Medina-Noyola}},\ }\href {\doibase 10.1103/PhysRevLett.75.759} {\bibfield
  {journal} {\bibinfo  {journal} {Phys. Rev. Lett.}\ }\textbf {\bibinfo
  {volume} {75}},\ \bibinfo {pages} {759} (\bibinfo {year}
  {1995}{\natexlab{a}})}\BibitemShut {NoStop}%
\bibitem [{\citenamefont {Viramontes-Gamboa}\ \emph
  {et~al.}(1995{\natexlab{b}})\citenamefont {Viramontes-Gamboa}, \citenamefont
  {Medina-Noyola},\ and\ \citenamefont {Arauz-Lara}}]{VirMedAra95PRE}%
  \BibitemOpen
  \bibfield  {author} {\bibinfo {author} {\bibfnamefont {G.}~\bibnamefont
  {Viramontes-Gamboa}}, \bibinfo {author} {\bibfnamefont {M.}~\bibnamefont
  {Medina-Noyola}}, \ and\ \bibinfo {author} {\bibfnamefont {J.~L.}\
  \bibnamefont {Arauz-Lara}},\ }\href {\doibase 10.1103/PhysRevE.52.4035}
  {\bibfield  {journal} {\bibinfo  {journal} {Phys. Rev. E}\ }\textbf {\bibinfo
  {volume} {52}},\ \bibinfo {pages} {4035} (\bibinfo {year}
  {1995}{\natexlab{b}})}\BibitemShut {NoStop}%
\bibitem [{\citenamefont {Scheidler}\ \emph
  {et~al.}(2002{\natexlab{a}})\citenamefont {Scheidler}, \citenamefont {Kob},
  \citenamefont {Binder},\ and\ \citenamefont {Parisi}}]{SchKobBinPar02PMB}%
  \BibitemOpen
  \bibfield  {author} {\bibinfo {author} {\bibfnamefont {P.}~\bibnamefont
  {Scheidler}}, \bibinfo {author} {\bibfnamefont {W.}~\bibnamefont {Kob}},
  \bibinfo {author} {\bibfnamefont {K.}~\bibnamefont {Binder}}, \ and\ \bibinfo
  {author} {\bibfnamefont {G.}~\bibnamefont {Parisi}},\ }\href {\doibase
  10.1080/13642810208221307} {\bibfield  {journal} {\bibinfo  {journal} {Phil.
  Mag. B}\ }\textbf {\bibinfo {volume} {82}},\ \bibinfo {pages} {283} (\bibinfo
  {year} {2002}{\natexlab{a}})}\BibitemShut {NoStop}%
\bibitem [{\citenamefont {Scheidler}\ \emph
  {et~al.}(2002{\natexlab{b}})\citenamefont {Scheidler}, \citenamefont {Kob},\
  and\ \citenamefont {Binder}}]{SchKobBin02EL}%
  \BibitemOpen
  \bibfield  {author} {\bibinfo {author} {\bibfnamefont {P.}~\bibnamefont
  {Scheidler}}, \bibinfo {author} {\bibfnamefont {W.}~\bibnamefont {Kob}}, \
  and\ \bibinfo {author} {\bibfnamefont {K.}~\bibnamefont {Binder}},\ }\href
  {\doibase 10.1209/epl/i2002-00182-9} {\bibfield  {journal} {\bibinfo
  {journal} {Europhys. Lett.}\ }\textbf {\bibinfo {volume} {59}},\ \bibinfo
  {pages} {701} (\bibinfo {year} {2002}{\natexlab{b}})}\BibitemShut {NoStop}%
\bibitem [{\citenamefont {Kim}(2003)}]{Kim03EL}%
  \BibitemOpen
  \bibfield  {author} {\bibinfo {author} {\bibfnamefont {K.}~\bibnamefont
  {Kim}},\ }\href {\doibase 10.1209/epl/i2003-00303-0} {\bibfield  {journal}
  {\bibinfo  {journal} {Europhys. Lett.}\ }\textbf {\bibinfo {volume} {61}},\
  \bibinfo {pages} {790} (\bibinfo {year} {2003})}\BibitemShut {NoStop}%
\bibitem [{\citenamefont {Fenz}\ \emph {et~al.}(2009)\citenamefont {Fenz},
  \citenamefont {Mryglod}, \citenamefont {Prytula},\ and\ \citenamefont
  {Folk}}]{FenMryPryFol09PRE}%
  \BibitemOpen
  \bibfield  {author} {\bibinfo {author} {\bibfnamefont {W.}~\bibnamefont
  {Fenz}}, \bibinfo {author} {\bibfnamefont {I.~M.}\ \bibnamefont {Mryglod}},
  \bibinfo {author} {\bibfnamefont {O.}~\bibnamefont {Prytula}}, \ and\
  \bibinfo {author} {\bibfnamefont {R.}~\bibnamefont {Folk}},\ }\href {\doibase
  10.1103/PhysRevE.80.021202} {\bibfield  {journal} {\bibinfo  {journal} {Phys.
  Rev. E}\ }\textbf {\bibinfo {volume} {80}},\ \bibinfo {pages} {021202}
  (\bibinfo {year} {2009})}\BibitemShut {NoStop}%
\bibitem [{\citenamefont {Chang}\ \emph {et~al.}(2004)\citenamefont {Chang},
  \citenamefont {Jagannathan},\ and\ \citenamefont
  {Yethiraj}}]{ChaJagYet04PRE}%
  \BibitemOpen
  \bibfield  {author} {\bibinfo {author} {\bibfnamefont {R.}~\bibnamefont
  {Chang}}, \bibinfo {author} {\bibfnamefont {K.}~\bibnamefont {Jagannathan}},
  \ and\ \bibinfo {author} {\bibfnamefont {A.}~\bibnamefont {Yethiraj}},\
  }\href {\doibase 10.1103/PhysRevE.69.051101} {\bibfield  {journal} {\bibinfo
  {journal} {Phys. Rev. E}\ }\textbf {\bibinfo {volume} {69}},\ \bibinfo
  {pages} {051101} (\bibinfo {year} {2004})}\BibitemShut {NoStop}%
\bibitem [{\citenamefont {Mittal}\ \emph {et~al.}(2006)\citenamefont {Mittal},
  \citenamefont {Errington},\ and\ \citenamefont {Truskett}}]{MitErrTru06PRE}%
  \BibitemOpen
  \bibfield  {author} {\bibinfo {author} {\bibfnamefont {J.}~\bibnamefont
  {Mittal}}, \bibinfo {author} {\bibfnamefont {J.~R.}\ \bibnamefont
  {Errington}}, \ and\ \bibinfo {author} {\bibfnamefont {T.~M.}\ \bibnamefont
  {Truskett}},\ }\href {\doibase 10.1103/PhysRevE.74.040102} {\bibfield
  {journal} {\bibinfo  {journal} {Phys. Rev. E}\ }\textbf {\bibinfo {volume}
  {74}},\ \bibinfo {pages} {040102} (\bibinfo {year} {2006})}\BibitemShut
  {NoStop}%
\bibitem [{\citenamefont {Ch{\'a}vez-Rojo}\ \emph {et~al.}(2008)\citenamefont
  {Ch{\'a}vez-Rojo}, \citenamefont {Ju{\'a}rez-Maldonado},\ and\ \citenamefont
  {Medina-Noyola}}]{ChaJuaMed08PRE}%
  \BibitemOpen
  \bibfield  {author} {\bibinfo {author} {\bibfnamefont {M.~A.}\ \bibnamefont
  {Ch{\'a}vez-Rojo}}, \bibinfo {author} {\bibfnamefont {R.}~\bibnamefont
  {Ju{\'a}rez-Maldonado}}, \ and\ \bibinfo {author} {\bibfnamefont
  {M.}~\bibnamefont {Medina-Noyola}},\ }\href {\doibase
  10.1103/PhysRevE.77.040401} {\bibfield  {journal} {\bibinfo  {journal} {Phys.
  Rev. E}\ }\textbf {\bibinfo {volume} {77}},\ \bibinfo {pages} {040401}
  (\bibinfo {year} {2008})}\BibitemShut {NoStop}%
\bibitem [{\citenamefont {Kim}\ \emph {et~al.}(2009)\citenamefont {Kim},
  \citenamefont {Miyazaki},\ and\ \citenamefont {Saito}}]{KimMiySai09EL}%
  \BibitemOpen
  \bibfield  {author} {\bibinfo {author} {\bibfnamefont {K.}~\bibnamefont
  {Kim}}, \bibinfo {author} {\bibfnamefont {K.}~\bibnamefont {Miyazaki}}, \
  and\ \bibinfo {author} {\bibfnamefont {S.}~\bibnamefont {Saito}},\ }\href
  {\doibase 10.1209/0295-5075/88/36002} {\bibfield  {journal} {\bibinfo
  {journal} {EPL}\ }\textbf {\bibinfo {volume} {88}},\ \bibinfo {pages} {36002}
  (\bibinfo {year} {2009})}\BibitemShut {NoStop}%
\bibitem [{\citenamefont {Kim}\ \emph {et~al.}(2010)\citenamefont {Kim},
  \citenamefont {Miyazaki},\ and\ \citenamefont {Saito}}]{KimMiySai10EPJST}%
  \BibitemOpen
  \bibfield  {author} {\bibinfo {author} {\bibfnamefont {K.}~\bibnamefont
  {Kim}}, \bibinfo {author} {\bibfnamefont {K.}~\bibnamefont {Miyazaki}}, \
  and\ \bibinfo {author} {\bibfnamefont {S.}~\bibnamefont {Saito}},\ }\href
  {\doibase 10.1140/epjst/e2010-01315-y} {\bibfield  {journal} {\bibinfo
  {journal} {Eur. Phys. J. Special Topics}\ }\textbf {\bibinfo {volume}
  {189}},\ \bibinfo {pages} {135} (\bibinfo {year} {2010})}\BibitemShut
  {NoStop}%
\bibitem [{\citenamefont {Kim}\ \emph {et~al.}(2011)\citenamefont {Kim},
  \citenamefont {Miyazaki},\ and\ \citenamefont {Saito}}]{KimMiySai11JPCM}%
  \BibitemOpen
  \bibfield  {author} {\bibinfo {author} {\bibfnamefont {K.}~\bibnamefont
  {Kim}}, \bibinfo {author} {\bibfnamefont {K.}~\bibnamefont {Miyazaki}}, \
  and\ \bibinfo {author} {\bibfnamefont {S.}~\bibnamefont {Saito}},\ }\href
  {\doibase 10.1088/0953-8984/23/23/234123} {\bibfield  {journal} {\bibinfo
  {journal} {J. Phys.: Condens. Matter}\ }\textbf {\bibinfo {volume} {23}},\
  \bibinfo {pages} {234123} (\bibinfo {year} {2011})}\BibitemShut {NoStop}%
\bibitem [{\citenamefont {Kurzidim}\ \emph {et~al.}(2010)\citenamefont
  {Kurzidim}, \citenamefont {Coslovich},\ and\ \citenamefont
  {Kahl}}]{KurCosKah10PRE}%
  \BibitemOpen
  \bibfield  {author} {\bibinfo {author} {\bibfnamefont {J.}~\bibnamefont
  {Kurzidim}}, \bibinfo {author} {\bibfnamefont {D.}~\bibnamefont {Coslovich}},
  \ and\ \bibinfo {author} {\bibfnamefont {G.}~\bibnamefont {Kahl}},\ }\href
  {\doibase 10.1103/PhysRevE.82.041505} {\bibfield  {journal} {\bibinfo
  {journal} {Phys. Rev. E}\ }\textbf {\bibinfo {volume} {82}},\ \bibinfo
  {pages} {041505} (\bibinfo {year} {2010})}\BibitemShut {NoStop}%
\bibitem [{\citenamefont {Bouchaud}\ and\ \citenamefont
  {Biroli}(2004)}]{BouBir04JCP}%
  \BibitemOpen
  \bibfield  {author} {\bibinfo {author} {\bibfnamefont {J.-P.}\ \bibnamefont
  {Bouchaud}}\ and\ \bibinfo {author} {\bibfnamefont {G.}~\bibnamefont
  {Biroli}},\ }\href {\doibase 10.1063/1.1796231} {\bibfield  {journal}
  {\bibinfo  {journal} {J. Chem. Phys.}\ }\textbf {\bibinfo {volume} {121}},\
  \bibinfo {pages} {7347} (\bibinfo {year} {2004})}\BibitemShut {NoStop}%
\bibitem [{\citenamefont {Montanari}\ and\ \citenamefont
  {Semerjian}(2006{\natexlab{a}})}]{MonSem06JSP_2}%
  \BibitemOpen
  \bibfield  {author} {\bibinfo {author} {\bibfnamefont {A.}~\bibnamefont
  {Montanari}}\ and\ \bibinfo {author} {\bibfnamefont {G.}~\bibnamefont
  {Semerjian}},\ }\href {\doibase 10.1007/s10955-006-9103-1} {\bibfield
  {journal} {\bibinfo  {journal} {J. Stat. Phys.}\ }\textbf {\bibinfo {volume}
  {124}},\ \bibinfo {pages} {103} (\bibinfo {year}
  {2006}{\natexlab{a}})}\BibitemShut {NoStop}%
\bibitem [{\citenamefont {M{\'e}zard}\ and\ \citenamefont
  {Montanari}(2006)}]{MezMon06JSP}%
  \BibitemOpen
  \bibfield  {author} {\bibinfo {author} {\bibfnamefont {M.}~\bibnamefont
  {M{\'e}zard}}\ and\ \bibinfo {author} {\bibfnamefont {A.}~\bibnamefont
  {Montanari}},\ }\href {\doibase 10.1007/s10955-006-9162-3} {\bibfield
  {journal} {\bibinfo  {journal} {J. Stat. Phys.}\ }\textbf {\bibinfo {volume}
  {124}},\ \bibinfo {pages} {1317} (\bibinfo {year} {2006})}\BibitemShut
  {NoStop}%
\bibitem [{\citenamefont {Montanari}\ and\ \citenamefont
  {Semerjian}(2006{\natexlab{b}})}]{MonSem06JSP}%
  \BibitemOpen
  \bibfield  {author} {\bibinfo {author} {\bibfnamefont {A.}~\bibnamefont
  {Montanari}}\ and\ \bibinfo {author} {\bibfnamefont {G.}~\bibnamefont
  {Semerjian}},\ }\href {\doibase 10.1007/s10955-006-9175-y} {\bibfield
  {journal} {\bibinfo  {journal} {J. Stat. Phys.}\ }\textbf {\bibinfo {volume}
  {125}},\ \bibinfo {pages} {23} (\bibinfo {year}
  {2006}{\natexlab{b}})}\BibitemShut {NoStop}%
\bibitem [{\citenamefont {Franz}\ and\ \citenamefont
  {Montanari}(2007)}]{FraMon07JPA}%
  \BibitemOpen
  \bibfield  {author} {\bibinfo {author} {\bibfnamefont {S.}~\bibnamefont
  {Franz}}\ and\ \bibinfo {author} {\bibfnamefont {A.}~\bibnamefont
  {Montanari}},\ }\href {\doibase 10.1088/1751-8113/40/11/F01} {\bibfield
  {journal} {\bibinfo  {journal} {J. Phys. A: Math. Gen.}\ }\textbf {\bibinfo
  {volume} {40}},\ \bibinfo {pages} {F251} (\bibinfo {year}
  {2007})}\BibitemShut {NoStop}%
\bibitem [{\citenamefont {Cavagna}\ \emph {et~al.}(2007)\citenamefont
  {Cavagna}, \citenamefont {Grigera},\ and\ \citenamefont
  {Verrocchio}}]{CavGriVer07PRL}%
  \BibitemOpen
  \bibfield  {author} {\bibinfo {author} {\bibfnamefont {A.}~\bibnamefont
  {Cavagna}}, \bibinfo {author} {\bibfnamefont {T.~S.}\ \bibnamefont
  {Grigera}}, \ and\ \bibinfo {author} {\bibfnamefont {P.}~\bibnamefont
  {Verrocchio}},\ }\href {\doibase 10.1103/PhysRevLett.98.187801} {\bibfield
  {journal} {\bibinfo  {journal} {Phys. Rev. Lett.}\ }\textbf {\bibinfo
  {volume} {98}},\ \bibinfo {pages} {187801} (\bibinfo {year}
  {2007})}\BibitemShut {NoStop}%
\bibitem [{\citenamefont {Biroli}\ \emph {et~al.}(2008)\citenamefont {Biroli},
  \citenamefont {Bouchaud}, \citenamefont {Cavagna}, \citenamefont {Grigera},\
  and\ \citenamefont {Verrocchio}}]{BirBouCavGriVer08NatPhys}%
  \BibitemOpen
  \bibfield  {author} {\bibinfo {author} {\bibfnamefont {G.}~\bibnamefont
  {Biroli}}, \bibinfo {author} {\bibfnamefont {J.-P.}\ \bibnamefont
  {Bouchaud}}, \bibinfo {author} {\bibfnamefont {A.}~\bibnamefont {Cavagna}},
  \bibinfo {author} {\bibfnamefont {T.~S.}\ \bibnamefont {Grigera}}, \ and\
  \bibinfo {author} {\bibfnamefont {P.}~\bibnamefont {Verrocchio}},\ }\href
  {\doibase 10.1038/nphys1050} {\bibfield  {journal} {\bibinfo  {journal}
  {Nature Phys.}\ }\textbf {\bibinfo {volume} {4}},\ \bibinfo {pages} {771}
  (\bibinfo {year} {2008})}\BibitemShut {NoStop}%
\bibitem [{\citenamefont {Cavagna}\ \emph
  {et~al.}(2010{\natexlab{a}})\citenamefont {Cavagna}, \citenamefont
  {Grigera},\ and\ \citenamefont {Verrocchio}}]{CavGriVer10JSMTE}%
  \BibitemOpen
  \bibfield  {author} {\bibinfo {author} {\bibfnamefont {A.}~\bibnamefont
  {Cavagna}}, \bibinfo {author} {\bibfnamefont {T.~S.}\ \bibnamefont
  {Grigera}}, \ and\ \bibinfo {author} {\bibfnamefont {P.}~\bibnamefont
  {Verrocchio}},\ }\href {\doibase 10.1088/1742-5468/2010/10/P10001} {\bibfield
   {journal} {\bibinfo  {journal} {J. Stat. Mech.}\ ,\ \bibinfo {pages}
  {P10001}} (\bibinfo {year} {2010}{\natexlab{a}})}\BibitemShut {NoStop}%
\bibitem [{\citenamefont {Cavagna}\ \emph
  {et~al.}(2010{\natexlab{b}})\citenamefont {Cavagna}, \citenamefont
  {Grigera},\ and\ \citenamefont {Verrocchio}}]{CavGriVer10preprint}%
  \BibitemOpen
  \bibfield  {author} {\bibinfo {author} {\bibfnamefont {A.}~\bibnamefont
  {Cavagna}}, \bibinfo {author} {\bibfnamefont {T.~S.}\ \bibnamefont
  {Grigera}}, \ and\ \bibinfo {author} {\bibfnamefont {P.}~\bibnamefont
  {Verrocchio}},\ }\href@noop {} {\enquote {\bibinfo {title} {\emph{Dynamic
  relaxation of a liquid cavity under amorphous boundary conditions}},}\
  }\bibinfo {howpublished} {ar{X}iv:1006.3746} (\bibinfo {year}
  {2010}{\natexlab{b}})\BibitemShut {NoStop}%
\bibitem [{\citenamefont {Karmakar}\ and\ \citenamefont
  {Procaccia}(2011)}]{KarPro11preprint}%
  \BibitemOpen
  \bibfield  {author} {\bibinfo {author} {\bibfnamefont {S.}~\bibnamefont
  {Karmakar}}\ and\ \bibinfo {author} {\bibfnamefont {I.}~\bibnamefont
  {Procaccia}},\ }\href@noop {} {\enquote {\bibinfo {title} {\emph{Exposing the
  static scale of the glass transition by random pinning}},}\ }\bibinfo
  {howpublished} {ar{X}iv:1105.4053} (\bibinfo {year} {2011})\BibitemShut
  {NoStop}%
\bibitem [{\citenamefont {Berthier}\ and\ \citenamefont
  {Kob}(2011)}]{BerKob11preprint}%
  \BibitemOpen
  \bibfield  {author} {\bibinfo {author} {\bibfnamefont {L.}~\bibnamefont
  {Berthier}}\ and\ \bibinfo {author} {\bibfnamefont {W.}~\bibnamefont {Kob}},\
  }\href@noop {} {\enquote {\bibinfo {title} {\emph{Static point-to-set
  correlations in glass-forming liquids}},}\ }\bibinfo {howpublished}
  {ar{X}iv:1105.6203} (\bibinfo {year} {2011})\BibitemShut {NoStop}%
\bibitem [{\citenamefont {Cammarota}\ and\ \citenamefont
  {Biroli}(2011)}]{CamBir11preprint}%
  \BibitemOpen
  \bibfield  {author} {\bibinfo {author} {\bibfnamefont {C.}~\bibnamefont
  {Cammarota}}\ and\ \bibinfo {author} {\bibfnamefont {G.}~\bibnamefont
  {Biroli}},\ }\href@noop {} {\enquote {\bibinfo {title} {\emph{Ideal glass
  transitions by random pinning}},}\ }\bibinfo {howpublished}
  {ar{X}iv:1106.5513} (\bibinfo {year} {2011})\BibitemShut {NoStop}%
\bibitem [{\citenamefont {Kob}\ \emph {et~al.}(2011)\citenamefont {Kob},
  \citenamefont {Rold{\'a}n-Vargas},\ and\ \citenamefont
  {Berthier}}]{KobRolBer11preprint}%
  \BibitemOpen
  \bibfield  {author} {\bibinfo {author} {\bibfnamefont {W.}~\bibnamefont
  {Kob}}, \bibinfo {author} {\bibfnamefont {S.}~\bibnamefont
  {Rold{\'a}n-Vargas}}, \ and\ \bibinfo {author} {\bibfnamefont
  {L.}~\bibnamefont {Berthier}},\ }\href@noop {} {\enquote {\bibinfo {title}
  {\emph{Non-monotonic temperature evolution of dynamic correlations in
  glass-forming liquids}},}\ }\bibinfo {howpublished} {ar{X}iv:1107.3928}
  (\bibinfo {year} {2011})\BibitemShut {NoStop}%
\bibitem [{\citenamefont {Charbonneau}\ \emph
  {et~al.}(2011{\natexlab{a}})\citenamefont {Charbonneau}, \citenamefont
  {Charbonneau},\ and\ \citenamefont {Tarjus}}]{ChaChaTar11preprint}%
  \BibitemOpen
  \bibfield  {author} {\bibinfo {author} {\bibfnamefont {B.}~\bibnamefont
  {Charbonneau}}, \bibinfo {author} {\bibfnamefont {P.}~\bibnamefont
  {Charbonneau}}, \ and\ \bibinfo {author} {\bibfnamefont {G.}~\bibnamefont
  {Tarjus}},\ }\href@noop {} {\enquote {\bibinfo {title} {\emph{Geometrical
  frustration and static correlations in a simple glass former}},}\ }\bibinfo
  {howpublished} {ar{X}iv:1108.2494} (\bibinfo {year}
  {2011}{\natexlab{a}})\BibitemShut {NoStop}%
\bibitem [{\citenamefont {Cruz~de Le\'on}\ \emph {et~al.}(1998)\citenamefont
  {Cruz~de Le\'on}, \citenamefont {Saucedo-Solorio},\ and\ \citenamefont
  {Arauz-Lara}}]{CruSauAra98PRL}%
  \BibitemOpen
  \bibfield  {author} {\bibinfo {author} {\bibfnamefont {G.}~\bibnamefont
  {Cruz~de Le\'on}}, \bibinfo {author} {\bibfnamefont {J.~M.}\ \bibnamefont
  {Saucedo-Solorio}}, \ and\ \bibinfo {author} {\bibfnamefont {J.~L.}\
  \bibnamefont {Arauz-Lara}},\ }\href {\doibase 10.1103/PhysRevLett.81.1122}
  {\bibfield  {journal} {\bibinfo  {journal} {Phys. Rev. Lett.}\ }\textbf
  {\bibinfo {volume} {81}},\ \bibinfo {pages} {1122} (\bibinfo {year}
  {1998})}\BibitemShut {NoStop}%
\bibitem [{\citenamefont {Cruz~de Le\'on}\ and\ \citenamefont
  {Arauz-Lara}(1999)}]{CruAra99PRE}%
  \BibitemOpen
  \bibfield  {author} {\bibinfo {author} {\bibfnamefont {G.}~\bibnamefont
  {Cruz~de Le\'on}}\ and\ \bibinfo {author} {\bibfnamefont {J.~L.}\
  \bibnamefont {Arauz-Lara}},\ }\href {\doibase 10.1103/PhysRevE.59.4203}
  {\bibfield  {journal} {\bibinfo  {journal} {Phys. Rev. E}\ }\textbf {\bibinfo
  {volume} {59}},\ \bibinfo {pages} {4203} (\bibinfo {year}
  {1999})}\BibitemShut {NoStop}%
\bibitem [{\citenamefont {G{\"o}tze}(2009)}]{gotzebook}%
  \BibitemOpen
  \bibfield  {author} {\bibinfo {author} {\bibfnamefont {W.}~\bibnamefont
  {G{\"o}tze}},\ }\href@noop {} {\emph {\bibinfo {title} {Complex Dynamics of
  Glass-Forming Liquids--A Mode-Coupling Theory}}}\ (\bibinfo  {publisher}
  {Oxford University},\ \bibinfo {address} {Oxford},\ \bibinfo {year}
  {2009})\BibitemShut {NoStop}%
\bibitem [{\citenamefont {Krakoviack}(2005{\natexlab{a}})}]{Kra05PRL}%
  \BibitemOpen
  \bibfield  {author} {\bibinfo {author} {\bibfnamefont {V.}~\bibnamefont
  {Krakoviack}},\ }\href {\doibase 10.1103/PhysRevLett.94.065703} {\bibfield
  {journal} {\bibinfo  {journal} {Phys. Rev. Lett.}\ }\textbf {\bibinfo
  {volume} {94}},\ \bibinfo {pages} {065703} (\bibinfo {year}
  {2005}{\natexlab{a}})}\BibitemShut {NoStop}%
\bibitem [{\citenamefont {Krakoviack}(2005{\natexlab{b}})}]{Kra05JPCM}%
  \BibitemOpen
  \bibfield  {author} {\bibinfo {author} {\bibfnamefont {V.}~\bibnamefont
  {Krakoviack}},\ }\href {\doibase 10.1088/0953-8984/17/45/049} {\bibfield
  {journal} {\bibinfo  {journal} {J. Phys.: Condens. Matter}\ }\textbf
  {\bibinfo {volume} {17}},\ \bibinfo {pages} {S3565} (\bibinfo {year}
  {2005}{\natexlab{b}})}\BibitemShut {NoStop}%
\bibitem [{\citenamefont {Krakoviack}(2007)}]{Kra07PRE}%
  \BibitemOpen
  \bibfield  {author} {\bibinfo {author} {\bibfnamefont {V.}~\bibnamefont
  {Krakoviack}},\ }\href {\doibase 10.1103/PhysRevE.75.031503} {\bibfield
  {journal} {\bibinfo  {journal} {Phys. Rev. E}\ }\textbf {\bibinfo {volume}
  {75}},\ \bibinfo {pages} {031503} (\bibinfo {year} {2007})}\BibitemShut
  {NoStop}%
\bibitem [{\citenamefont {Krakoviack}(2009)}]{Kra09PRE}%
  \BibitemOpen
  \bibfield  {author} {\bibinfo {author} {\bibfnamefont {V.}~\bibnamefont
  {Krakoviack}},\ }\href {\doibase 10.1103/PhysRevE.79.061501} {\bibfield
  {journal} {\bibinfo  {journal} {Phys. Rev. E}\ }\textbf {\bibinfo {volume}
  {79}},\ \bibinfo {pages} {061501} (\bibinfo {year} {2009})}\BibitemShut
  {NoStop}%
\bibitem [{\citenamefont {Kurzidim}\ \emph {et~al.}(2009)\citenamefont
  {Kurzidim}, \citenamefont {Coslovich},\ and\ \citenamefont
  {Kahl}}]{KurCosKah09PRL}%
  \BibitemOpen
  \bibfield  {author} {\bibinfo {author} {\bibfnamefont {J.}~\bibnamefont
  {Kurzidim}}, \bibinfo {author} {\bibfnamefont {D.}~\bibnamefont {Coslovich}},
  \ and\ \bibinfo {author} {\bibfnamefont {G.}~\bibnamefont {Kahl}},\ }\href
  {\doibase 10.1103/PhysRevLett.103.138303} {\bibfield  {journal} {\bibinfo
  {journal} {Phys. Rev. Lett.}\ }\textbf {\bibinfo {volume} {103}},\ \bibinfo
  {pages} {138303} (\bibinfo {year} {2009})}\BibitemShut {NoStop}%
\bibitem [{\citenamefont {Kurzidim}\ \emph {et~al.}(2011)\citenamefont
  {Kurzidim}, \citenamefont {Coslovich},\ and\ \citenamefont
  {Kahl}}]{KurCosKah11JPCM}%
  \BibitemOpen
  \bibfield  {author} {\bibinfo {author} {\bibfnamefont {J.}~\bibnamefont
  {Kurzidim}}, \bibinfo {author} {\bibfnamefont {D.}~\bibnamefont {Coslovich}},
  \ and\ \bibinfo {author} {\bibfnamefont {G.}~\bibnamefont {Kahl}},\ }\href
  {\doibase 10.1088/0953-8984/23/23/234122} {\bibfield  {journal} {\bibinfo
  {journal} {J. Phys.: Condens. Matter}\ }\textbf {\bibinfo {volume} {23}},\
  \bibinfo {pages} {234122} (\bibinfo {year} {2011})}\BibitemShut {NoStop}%
\bibitem [{\citenamefont {Kirkpatrick}\ \emph {et~al.}(1989)\citenamefont
  {Kirkpatrick}, \citenamefont {Thirumalai},\ and\ \citenamefont
  {Wolynes}}]{KirThiWol89PRA}%
  \BibitemOpen
  \bibfield  {author} {\bibinfo {author} {\bibfnamefont {T.~R.}\ \bibnamefont
  {Kirkpatrick}}, \bibinfo {author} {\bibfnamefont {D.}~\bibnamefont
  {Thirumalai}}, \ and\ \bibinfo {author} {\bibfnamefont {P.~G.}\ \bibnamefont
  {Wolynes}},\ }\href {\doibase 10.1103/PhysRevA.40.1045} {\bibfield  {journal}
  {\bibinfo  {journal} {Phys. Rev. A}\ }\textbf {\bibinfo {volume} {40}},\
  \bibinfo {pages} {1045} (\bibinfo {year} {1989})}\BibitemShut {NoStop}%
\bibitem [{\citenamefont {Lubchenko}\ and\ \citenamefont
  {Wolynes}(2007)}]{LubWol07ARPC}%
  \BibitemOpen
  \bibfield  {author} {\bibinfo {author} {\bibfnamefont {V.}~\bibnamefont
  {Lubchenko}}\ and\ \bibinfo {author} {\bibfnamefont {P.~G.}\ \bibnamefont
  {Wolynes}},\ }\href {\doibase 10.1146/annurev.physchem.58.032806.104653}
  {\bibfield  {journal} {\bibinfo  {journal} {Ann. Rev. Phys. Chem.}\ }\textbf
  {\bibinfo {volume} {58}},\ \bibinfo {pages} {235} (\bibinfo {year}
  {2007})}\BibitemShut {NoStop}%
\bibitem [{\citenamefont {Schmid}\ and\ \citenamefont
  {Schilling}(2010)}]{SchSch10PRE}%
  \BibitemOpen
  \bibfield  {author} {\bibinfo {author} {\bibfnamefont {B.}~\bibnamefont
  {Schmid}}\ and\ \bibinfo {author} {\bibfnamefont {R.}~\bibnamefont
  {Schilling}},\ }\href {\doibase 10.1103/PhysRevE.81.041502} {\bibfield
  {journal} {\bibinfo  {journal} {Phys. Rev. E}\ }\textbf {\bibinfo {volume}
  {81}},\ \bibinfo {pages} {041502} (\bibinfo {year} {2010})}\BibitemShut
  {NoStop}%
\bibitem [{\citenamefont {Ikeda}\ and\ \citenamefont
  {Miyazaki}(2010)}]{IkeMiy10PRL}%
  \BibitemOpen
  \bibfield  {author} {\bibinfo {author} {\bibfnamefont {A.}~\bibnamefont
  {Ikeda}}\ and\ \bibinfo {author} {\bibfnamefont {K.}~\bibnamefont
  {Miyazaki}},\ }\href {\doibase 10.1103/PhysRevLett.104.255704} {\bibfield
  {journal} {\bibinfo  {journal} {Phys. Rev. Lett.}\ }\textbf {\bibinfo
  {volume} {104}},\ \bibinfo {pages} {255704} (\bibinfo {year}
  {2010})}\BibitemShut {NoStop}%
\bibitem [{\citenamefont {Charbonneau}\ \emph
  {et~al.}(2011{\natexlab{b}})\citenamefont {Charbonneau}, \citenamefont
  {Ikeda}, \citenamefont {Parisi},\ and\ \citenamefont
  {Zamponi}}]{ChaIkeParZam11preprint}%
  \BibitemOpen
  \bibfield  {author} {\bibinfo {author} {\bibfnamefont {P.}~\bibnamefont
  {Charbonneau}}, \bibinfo {author} {\bibfnamefont {A.}~\bibnamefont {Ikeda}},
  \bibinfo {author} {\bibfnamefont {G.}~\bibnamefont {Parisi}}, \ and\ \bibinfo
  {author} {\bibfnamefont {F.}~\bibnamefont {Zamponi}},\ }\href@noop {}
  {\enquote {\bibinfo {title} {\emph{Glass transition and random close packing
  in 3+ dimensions}},}\ }\bibinfo {howpublished} {ar{X}iv:1107.4666} (\bibinfo
  {year} {2011}{\natexlab{b}})\BibitemShut {NoStop}%
\bibitem [{\citenamefont {Given}\ and\ \citenamefont
  {Stell}(1992)}]{GivSte92JCP}%
  \BibitemOpen
  \bibfield  {author} {\bibinfo {author} {\bibfnamefont {J.~A.}\ \bibnamefont
  {Given}}\ and\ \bibinfo {author} {\bibfnamefont {G.}~\bibnamefont {Stell}},\
  }\href {\doibase 10.1063/1.463883} {\bibfield  {journal} {\bibinfo  {journal}
  {J. Chem. Phys.}\ }\textbf {\bibinfo {volume} {97}},\ \bibinfo {pages} {4573}
  (\bibinfo {year} {1992})}\BibitemShut {NoStop}%
\bibitem [{\citenamefont {Lomba}\ \emph {et~al.}(1993)\citenamefont {Lomba},
  \citenamefont {Given}, \citenamefont {Stell}, \citenamefont {Weis},\ and\
  \citenamefont {Levesque}}]{LomGivSteWeiLev93PRE}%
  \BibitemOpen
  \bibfield  {author} {\bibinfo {author} {\bibfnamefont {E.}~\bibnamefont
  {Lomba}}, \bibinfo {author} {\bibfnamefont {J.~A.}\ \bibnamefont {Given}},
  \bibinfo {author} {\bibfnamefont {G.}~\bibnamefont {Stell}}, \bibinfo
  {author} {\bibfnamefont {J.~J.}\ \bibnamefont {Weis}}, \ and\ \bibinfo
  {author} {\bibfnamefont {D.}~\bibnamefont {Levesque}},\ }\href {\doibase
  10.1103/PhysRevE.48.233} {\bibfield  {journal} {\bibinfo  {journal} {Phys.
  Rev. E}\ }\textbf {\bibinfo {volume} {48}},\ \bibinfo {pages} {233} (\bibinfo
  {year} {1993})}\BibitemShut {NoStop}%
\bibitem [{\citenamefont {Given}\ and\ \citenamefont
  {Stell}(1994)}]{GivSte94PA}%
  \BibitemOpen
  \bibfield  {author} {\bibinfo {author} {\bibfnamefont {J.~A.}\ \bibnamefont
  {Given}}\ and\ \bibinfo {author} {\bibfnamefont {G.}~\bibnamefont {Stell}},\
  }\href {\doibase 10.1016/0378-4371(94)90200-3} {\bibfield  {journal}
  {\bibinfo  {journal} {Physica A}\ }\textbf {\bibinfo {volume} {209}},\
  \bibinfo {pages} {495} (\bibinfo {year} {1994})}\BibitemShut {NoStop}%
\bibitem [{\citenamefont {Rosinberg}\ \emph {et~al.}(1994)\citenamefont
  {Rosinberg}, \citenamefont {Tarjus},\ and\ \citenamefont
  {Stell}}]{RosTarSte94JCP}%
  \BibitemOpen
  \bibfield  {author} {\bibinfo {author} {\bibfnamefont {M.~L.}\ \bibnamefont
  {Rosinberg}}, \bibinfo {author} {\bibfnamefont {G.}~\bibnamefont {Tarjus}}, \
  and\ \bibinfo {author} {\bibfnamefont {G.}~\bibnamefont {Stell}},\ }\href
  {\doibase 10.1063/1.467182} {\bibfield  {journal} {\bibinfo  {journal} {J.
  Chem. Phys.}\ }\textbf {\bibinfo {volume} {100}},\ \bibinfo {pages} {5172}
  (\bibinfo {year} {1994})}\BibitemShut {NoStop}%
\bibitem [{\citenamefont {Lifshits}\ \emph {et~al.}(1988)\citenamefont
  {Lifshits}, \citenamefont {Gredeskul},\ and\ \citenamefont
  {Pastur}}]{LifGrePasbook}%
  \BibitemOpen
  \bibfield  {author} {\bibinfo {author} {\bibfnamefont {I.~M.}\ \bibnamefont
  {Lifshits}}, \bibinfo {author} {\bibfnamefont {S.~A.}\ \bibnamefont
  {Gredeskul}}, \ and\ \bibinfo {author} {\bibfnamefont {L.~A.}\ \bibnamefont
  {Pastur}},\ }\href@noop {} {\emph {\bibinfo {title} {Introduction to the
  theory of disordered systems}}}\ (\bibinfo  {publisher} {Wiley},\ \bibinfo
  {address} {New York},\ \bibinfo {year} {1988})\BibitemShut {NoStop}%
\bibitem [{\citenamefont {Ford}\ and\ \citenamefont
  {Glandt}(1994)}]{ForGla94JCP}%
  \BibitemOpen
  \bibfield  {author} {\bibinfo {author} {\bibfnamefont {D.~M.}\ \bibnamefont
  {Ford}}\ and\ \bibinfo {author} {\bibfnamefont {E.~D.}\ \bibnamefont
  {Glandt}},\ }\href {\doibase 10.1063/1.466485} {\bibfield  {journal}
  {\bibinfo  {journal} {J. Chem. Phys.}\ }\textbf {\bibinfo {volume} {100}},\
  \bibinfo {pages} {2391} (\bibinfo {year} {1994})}\BibitemShut {NoStop}%
\bibitem [{\citenamefont {Meroni}\ \emph {et~al.}(1996)\citenamefont {Meroni},
  \citenamefont {Levesque},\ and\ \citenamefont {Weis}}]{MerLevWei96JCP}%
  \BibitemOpen
  \bibfield  {author} {\bibinfo {author} {\bibfnamefont {A.}~\bibnamefont
  {Meroni}}, \bibinfo {author} {\bibfnamefont {D.}~\bibnamefont {Levesque}}, \
  and\ \bibinfo {author} {\bibfnamefont {J.~J.}\ \bibnamefont {Weis}},\ }\href
  {\doibase 10.1063/1.471954} {\bibfield  {journal} {\bibinfo  {journal} {J.
  Chem. Phys.}\ }\textbf {\bibinfo {volume} {105}},\ \bibinfo {pages} {1101}
  (\bibinfo {year} {1996})}\BibitemShut {NoStop}%
\bibitem [{Note1()}]{Note1}%
  \BibitemOpen
  \bibinfo {note} {In fact, the MCT does predict a re-entrant behavior for the
  diffusion-localization transition line of the QA system as well, but with an
  exceedingly small amplitude. The difficulties raised by this result and a
  possible physical interpretation are discussed in Ref.~\cite
  {Kra09PRE}.}\BibitemShut {Stop}%
\end{thebibliography}
%../../Bibtex/confinedfluids_exp}

\end{document}